\documentclass[final,5p,times,twocolumn]{elsarticle}
\usepackage{graphics}

\journal{Astroparticle Physics}
\begin{document}

\begin{frontmatter}

\title{Performance of a large TeO$_2$ crystal as a cryogenic  bolometer in searching for neutrinoless double beta decay}

\author[INFNRM1,UNIROMA1]{L.~Cardani}
\author[INFNMIB,UNIMIB]{L.~Gironi}
\author[LBL]{J.W.~Beeman}
\author[INFNRM1]{I.~Dafinei}
\author[SICCAS]{Z.~Ge}
\author[INFNMIB]{G.~Pessina}
\author[INFNMIB]{S. Pirro \corref{cor1}}\ead{Stefano.Pirro@mib.infn.it}
\author[SICCAS]{Y.~Zhu}

\cortext[cor1]{Corresponding author}

\address[INFNRM1]{INFN - Sezione di Roma 1 I 00185 Roma - Italy}
\address[UNIROMA1]{Dipartimento di Fisica - Universit\`{a} di Roma La Sapienza I 00185 Roma - Italy}
\address[INFNMIB]{INFN - Sezione di Milano Bicocca I 20126 Milano - Italy}
\address[UNIMIB]{Dipartimento di Fisica - Universit\`{a} di Milano Bicocca I 20126 Milano - Italy}
\address[LBL]{Lawrence Berkeley National Laboratory , Berkeley, California 94720, USA}
\address[SICCAS]{Shanghai Institute of Ceramics, Chinese Academy of Science, Shanghai 200050, PR China}

\date{\today}

\begin{abstract}
Bolometers are  ideal devices in the search for neutrinoless Double Beta Decay (0$\nu$DBD). Enlarging the mass of 
individual  detectors would simplify   the construction of a large experiment, but would also decrease the background per unit mass  
induced  by $\alpha$-emitters located close to the surfaces and background arising from  external and internal $\gamma$'s.
We present the very promising results obtained with a 2.13 kg  TeO$_2$  crystal.   
This bolometer, cooled down to a temperature of 10.5 mK in a dilution refrigerator located deep underground in the Gran Sasso 
National Laboratories, represents  the largest thermal detector ever operated.
The detector exhibited  an energy resolution spanning a range from  3.9 keV  (at 145 keV) to 7.8 keV  (at the 
2615 $\gamma$-line of $^{208}$Tl) FWHM. We  discuss the decrease in the background per unit mass that can be achieved increasing the mass
of a bolometer. 
\end{abstract}


\end{frontmatter}

\section{Introduction}
\label{sec:Introduction}
Neutrinoless Double Beta Decay (0$\nu$DBD)   is a rare nuclear process hypothesized to occur in certain nuclei. If observed, it would give 
important information about the properties of the neutrino and the weak interaction. 
Double Beta Decay  searches \cite{DBDgeneral1,DBDgeneral2,DBDgeneral3} gained critical importance after the discovery of  
neutrino oscillations and many experiments are  concluding R\&D   and other are now under construction. 

Thermal bolometers are  ideal detectors for this survey because they can be composed by  most of the more interesting  $2\beta$-emitters and,
fundamental for next generation experiments, they show an excellent  energy resolution. 
The Cuoricino experiment \cite{Cuoricino-2011}, which constituted of  an array of 62 TeO$_2$ (750 g) crystal bolometers, demonstrated  the power of 
this technique and established the basis for the CUORE experiment \cite{Arnaboldi2004}, which will operate 988 TeO$_2$ crystals of the same size.
In addition to $^{130}$Te,   $2\beta$-scintillating  bolometers \cite{PHAN-2006} based on 
$^{116}$Cd \cite{Astro-CdWO4}, $^{100}$Mo \cite{JINST-ZnMoO4,NIMA-CaMoO4}, and $^{82}$Se \cite{Astro-ZnSe} 
were recently operated with success.

In such experiments, increasing the mass of the individual detector module can be extremely helpful, for several reasons.
First, it improves the Peak-to-Compton ratio for $\gamma$-ray interactions, enabling not only better identification of environmental background
but also  decreasing  the continuum induced by Compton and multi-Compton scattering. 
Second, the reduction in the surface-to-volume ratio reduces the background per unit mass from surface 
impurities \cite{fondoBB}. Moreover,  the total $2\beta$-efficiency (related to the full containment of the 2 e$^{-}$ emitted in the decay) 
of the detector will slightly increase. 
Last,  a  large-mass experiment inevitably requires  the use of an array of detectors, so a larger individual detector  corresponds to a lower number 
of readout channels, and a simpler setup.

\section{Growth of a 2.13 kg TeO$_2$ crystal }
\label{sec:Growth}

\begin{figure}
\begin{center}
\resizebox{0.48\textwidth}{!}
{\includegraphics{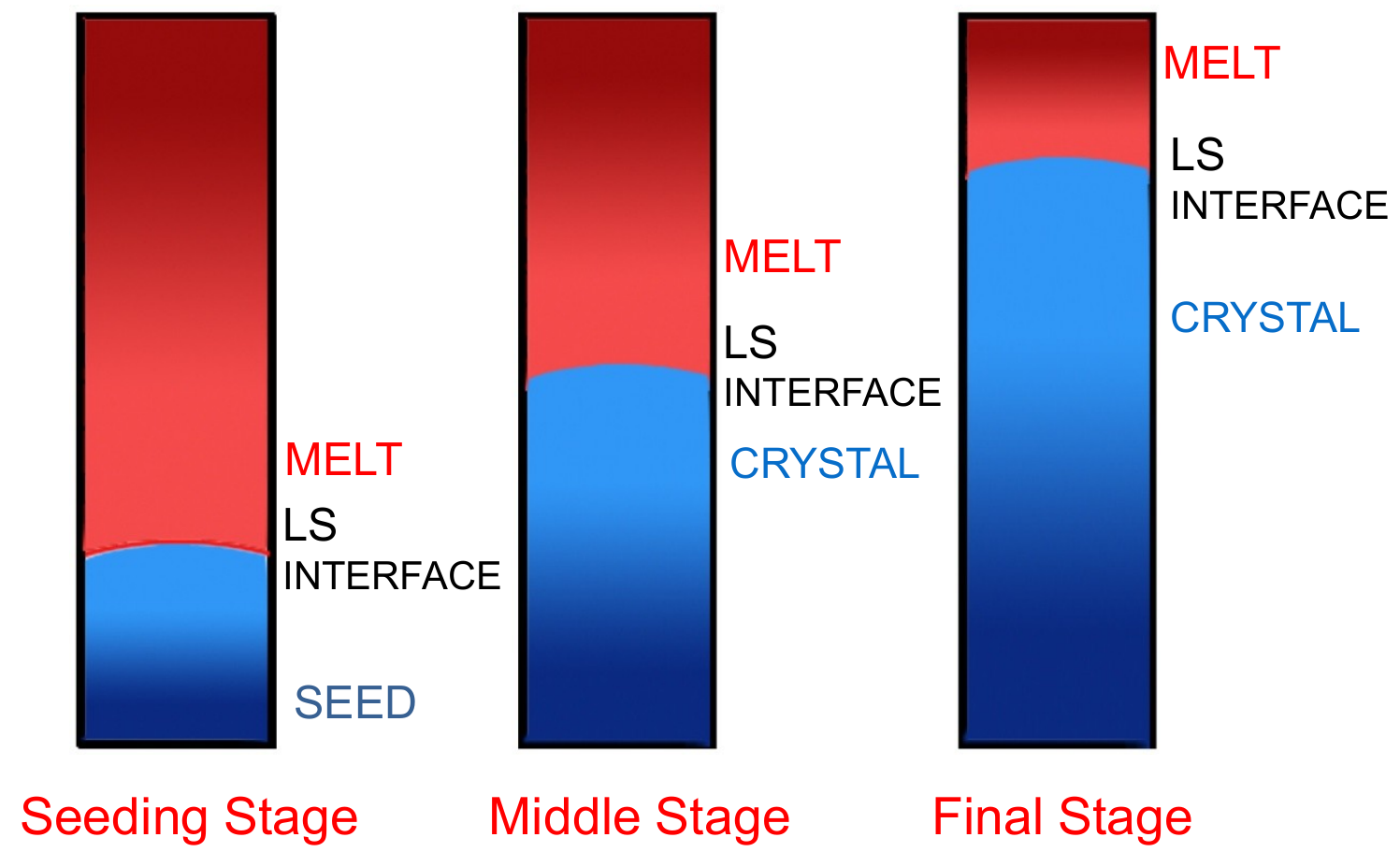}}
\caption{Sketch map of the growth stages in Bridgman method.}
\label{fig:fig0}      
\end{center}
\end{figure}

The TeO$_2$ crystal studied in this  work was grown using the modified Bridgman method, described in 
detail in previous articles \cite{JCG-2006,TeO2-crystal}. The main challenge in growing very large crystals using this technique  consists in maintaining 
an adequate thermal compensation during all stages of its growth, especially in the last (Fig.~\ref{fig:fig0}). 
Thermal compensation is needed to guarantee the flat or slightly convex liquidus-solidus (LS) interface essential to obtaining  
a good crystal perfection along the whole ingot. One of the peculiarities of our growth method is that the seed end 
of the crucible remains outside the furnace cavity (i.e. in open air) during all stages of growth, so controlling 
the LS interface is very difficult. Moreover, in the final stage almost the entire crucible is exposed to open air resulting in sizable thermal 
radiation with the need  of thermal compensation. In the particular case of growing a very large crystal the thermal 
compensation was even more critical and required implementing different thermal compensation rates during the growth of an 
approximately 2.5 kg TeO$_2$ ingot. In the final phase of cutting and polishing the as-grown ingot, a compromise was adopted in order to maximize 
the crystal weight  while maintaining a reasonable  shape and quality standards. In particular, two of the crystal corners were discarded, corresponding to 
roughly 2 cm$^3$. Some inclusions are present close to one of the crystal faces. The final crystal  shows a slightly truncated-pyramidal 
shape with a rectangular cross section. The dimensions of the boundary sections are 54.7 $\times$ 59.6 mm$^2$ and 54.0 $\times$ 58.2 mm$^2$. 
The length is 111.3 mm and its total  weight is 2.133 kg.

\section{Experimental details}
\label{sec:Experimental-details}
\begin{figure}
\begin{center}
\resizebox{0.48\textwidth}{!}
{\includegraphics{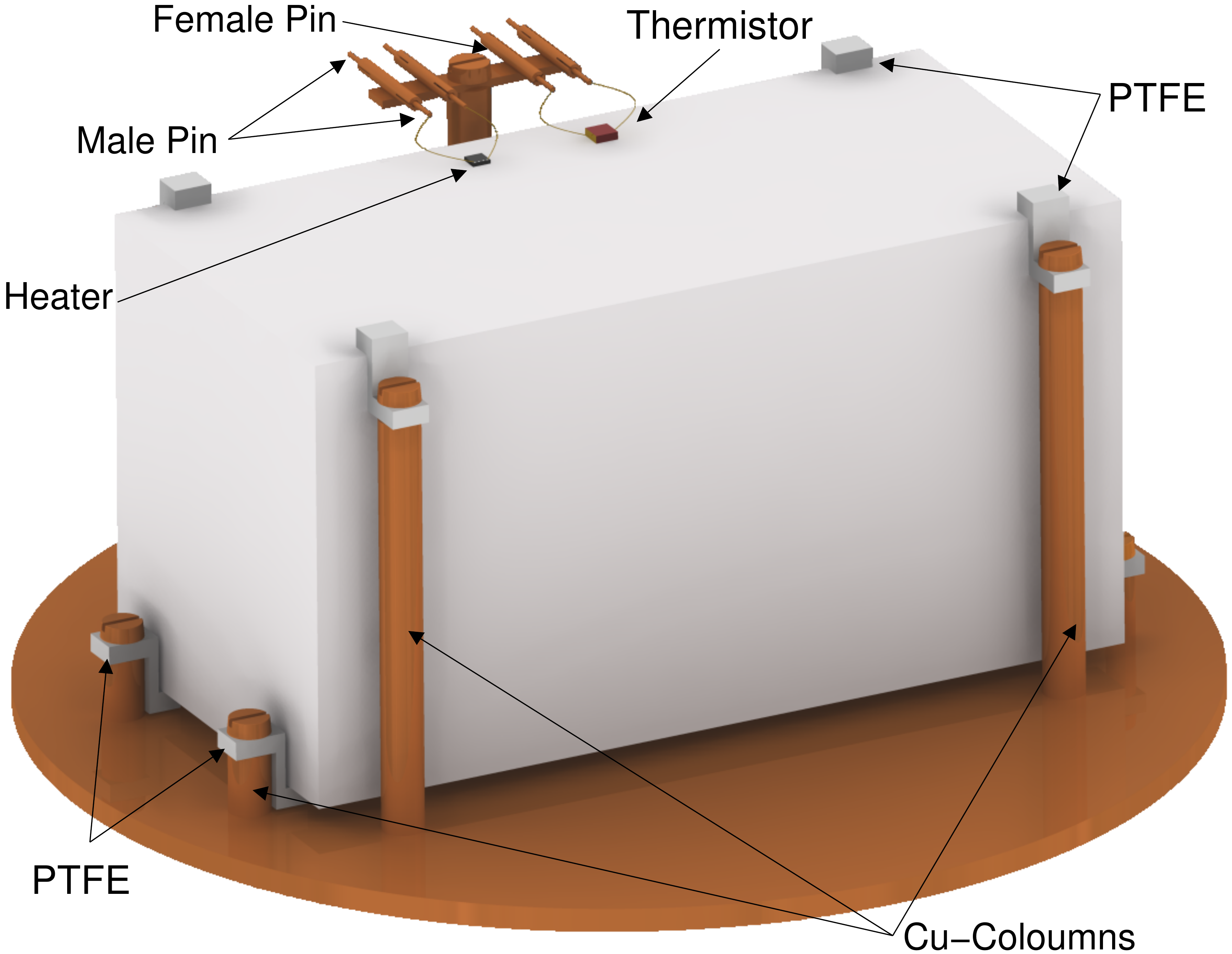}}
\caption{The detector setup. The crystal is held by means of eight S-shaped PTFE supports mounted on Cu columns. The thermistors and the heater contacts
are realized through crimped Cu pipes glued on a Cu plate. The entire setup is enclosed inside a Cu shield kept at base temperature.}
\label{fig:fig1}      
\end{center}
\end{figure}

The TeO$_2$ crystal bolometer  is secured by means of eight S-shaped PTFE supports mounted on Cu columns (Fig.~\ref{fig:fig1}); this is the simplest way to hold 
a crystal of non-standard shape. The S-shape of the Teflon supports   ensures that with the decrease of the temperature the crystal is clasped
tighter, due to the fact that the thermal contraction of PTFE is larger than TeO$_2$.
The temperature sensor  is a 3x3x1 mm$^3$ neutron transmutation doped Germanium thermistor, identical to the ones  used 
in the Cuoricino  experiment. It is thermally coupled to the crystal via 9 glue spots of $\sim$~0.6 mm diameter and $\sim$50 $\mu$m height.
In addition,  a $\sim$300 k$\Omega$ resistor made of  a heavily doped  meander on a 3.5 mm$^3$ silicon chip, is attached to  
each crystal and acts as a heater to  stabilize the gain of the bolometer \cite{ALES98,PESS-2003}. 
The 50 $\mu$m gold wires ball-bonded on thermistor and heater are crimped into  0.65 mm copper tubes (``male'' pin) inserted into larger copper tubes (``female'' pin) glued 
(electrically insulated) on a copper plate. Twisted constantan wires having a diameter  60 $\mu$m  (not shown in Fig.~\ref{fig:fig1}) are 
crimped in similar Cu tubes on the opposite end of the female connectors and carry the electrical signal up to the cryostat's  Mixing Chamber, where
a custom wiring brings the electrical signal  up to the  front-end electronics, located just outside the cryostat. 

The detector was operated deep underground in the Gran Sasso National Laboratories in the CUORE R\&D test cryostat. 
The details of the  the cryogenic facility and its electronics can be found elsewhere \cite{PESS-2002,NIMA-2004,NIMA-2006-B,NIMA-2006-C}.

Heat pulses, produced by  particle interactions in  the TeO$_2$ crystal are transduced into voltage pulses by 
the NTD thermistor, and are then  amplified and fed into a 16 bit NI 6225 USB ADC unit. 
The entire waveform (``raw pulse'')  of each triggered voltage pulse is sampled and acquired.
The amplitude  and the shape of the voltage pulse is then determined by an off line analysis which uses the 
Optimum Filter (O.F.) technique \cite{Arnaboldi2004,GattiManfredi-86}. The signal amplitude is computed as the maximum of the optimally 
filtered pulse, while the signal shape is evaluated on the basis of several different parameters: the rise time ($\tau_{rise}$) and 
the decay time ($\tau_{decay}$) are evaluated from the raw pulse as (t$_{90\%}$-t$_{10\%}$) and (t$_{30\%}$-t$_{90\%}$), respectively.

\section{Detector Performaces}
\label{sec:Detector_Performaces}

The detector was operated at a temperature of $\approx$ 10.5 mK. The corresponding working resistance of the thermistor was 65 M$\Omega$.
The main characteristics of the detector are reported in Tab.~\ref{Tab:tab1}.

\begin{table}[htb]
\begin{center}
\caption{Main parameters of the crystal bolometer. The second column represents the theoretical resolution given by the Optimum Filter.
The last column represents the absolute signal read out across the thermistor. The electronics has a six-pole Bessel filtering stage at 12 Hz.}
\vspace{.2 cm}
\label{Tab:tab1}
\begin{tabular}{@{}ccccc}
\hline
R  	  		& FWHM (O.F.) 		&  T$_{rise}$ 	  & T$_{decay}$    & Signal 				 \\

[M$\Omega$] & [keV] 			& 	[ms]  		  &	[ms] 		   & [$\mu$V/MeV]	         \\
\hline
65	  		& 	 3.7 		 	&   29	 		  & 95			   & 24				         \\
\hline
\end{tabular}
\end{center}
\end{table}

\begin{figure}
\begin{center}
\resizebox{0.48\textwidth}{!}
{\includegraphics{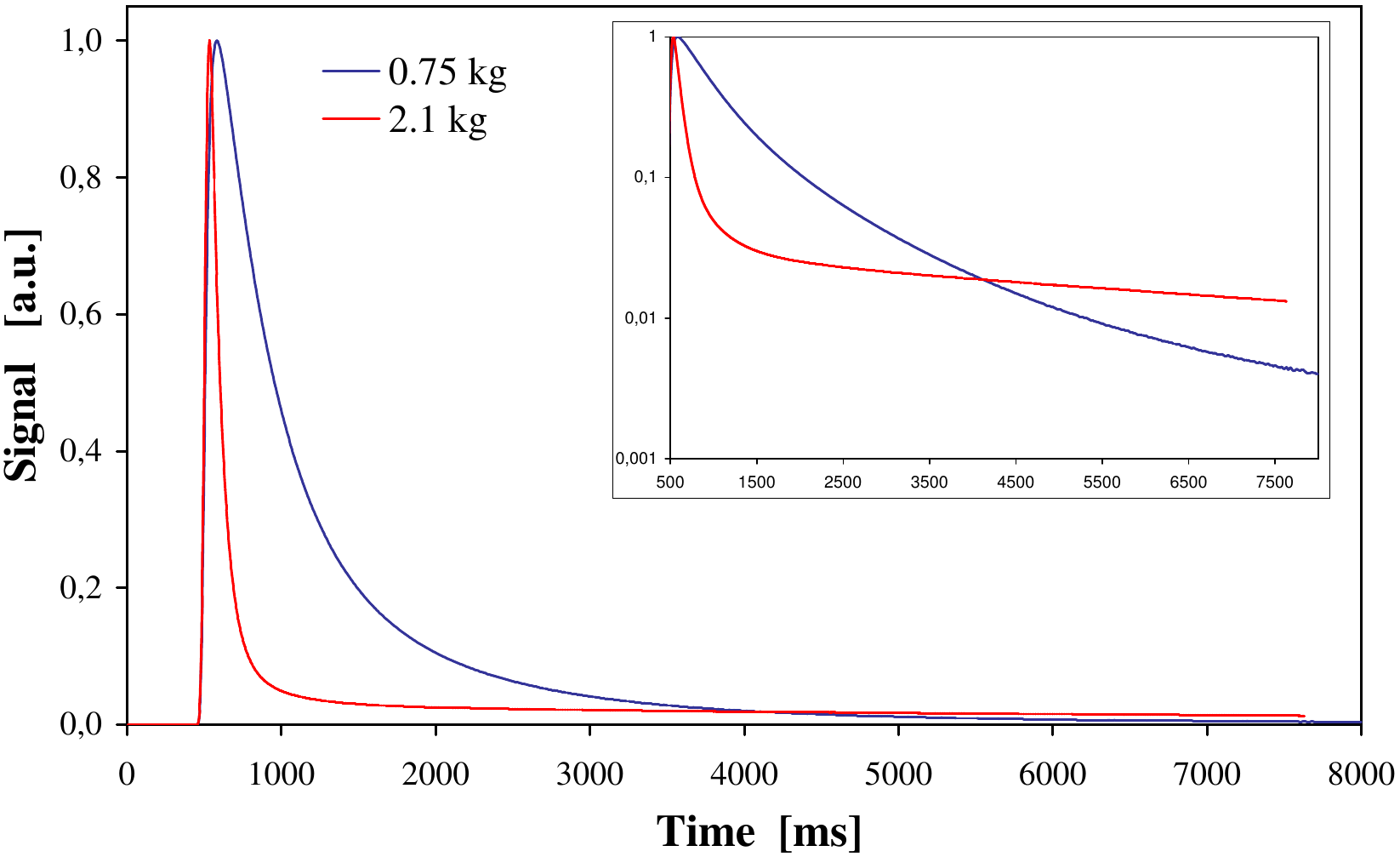}}
\caption{Average thermal pulse obtained for the  2.1 kg crystal under study (light red). For comparison  we also plot the average pulse obtained with  
750 g TeO$_2$ crystal (dark blue) under similar conditions (temperature and resistance). The signals are normalized to the rising edge. 
The inset shows  that the 2.1 kg crystal has a predominant fast decay  and a second, very slow decay constant.
Also the rise time of the large crystal results $\approx$ 2 times faster with respect to the smaller crystal.}
\label{fig:fig2}      
\end{center}
\end{figure}

The most remarkable feature of this large bolometer  is the signal shape. In particular,  the decay time shows two unexpected features: it 
abruptly decreases  to $\sim$ 8\% of the signal height and then  shows an extremely long decay constant. An example of  this behaviour is presented in 
Fig.~\ref{fig:fig2}, along with  the average thermal pulse\footnote{The average thermal pulse, i.e. the shape of a pulse 
in absence of noise, is computed from a proper average of a large number of raw pulses.} of a Cuoricino (750 g) TeO$_2$ crystal for comparison.

The mean signal decay time observed in the Cuoricino experiment was   of the order of $\sim$ 250 ms, while the mean rise time was of the order 
of $\sim$80 ms.
From a theoretical point of view the decay time for a perfect crystal is  given by 
$\tau$=C/G, where C  represents the heat capacity of the  crystal ($\propto$~mass) and G is the thermal link to the heat sink 
(dominated by the PTFE supports).
Given the fact that the PTFE supports used in Cuoricino were identical in shape and number, one would have expected a decay 
time of the order of $\approx$700 ms instead of 95 ms.
In fact this behaviour is  observed in crystal absorbers in which the crystal structure exhibits some defects. In the specific case of TeO$_2$ 
crystals, this behaviour was observed in the enriched ones (operated in Cuoricino) which, being the only ``imperfect'' TeO$_2$ tested up 
to now, show a pulse  shape very similar to the one observed in our large crystal under study.

The second characteristic that is normally associated with non perfect crystal lattices (often  resulting also in poly-crystalline structures) 
is a reduction in the absolute signal height. The observed value of 24 $\mu$V/MeV is rather small  when compared with the mean value of 
$\sim $150 $\mu$V/MeV observed in Cuoricino  at a similar working resistance (i.e. temperature)  \cite{NIMA-Pirro}.
We believe that this unusual thermal response can be definitely ascribed to the imperfections present in the crystal, as described in 
Sec.~\ref{sec:Growth}.
These imperfections normally imply  a degradation  in the energy resolution of the device as well. This can be qualitatively described in terms of 
position effects due to localized imperfections in which the interacting particle  could thermalize slightly differently. 
As  example, in Cuoricino the mean FWHM energy resolution  of the  enriched crystals (evaluated at 2615 keV) was  $\approx$ 15 keV, while the 
resolution of the same-sized (3$\times$3$\times$6 cm$^3$) natural crystal was $\approx$ 9 keV \cite{Cuoricino-2011}. 

\begin{figure}
\begin{center}
\resizebox{0.48\textwidth}{!}
{\includegraphics{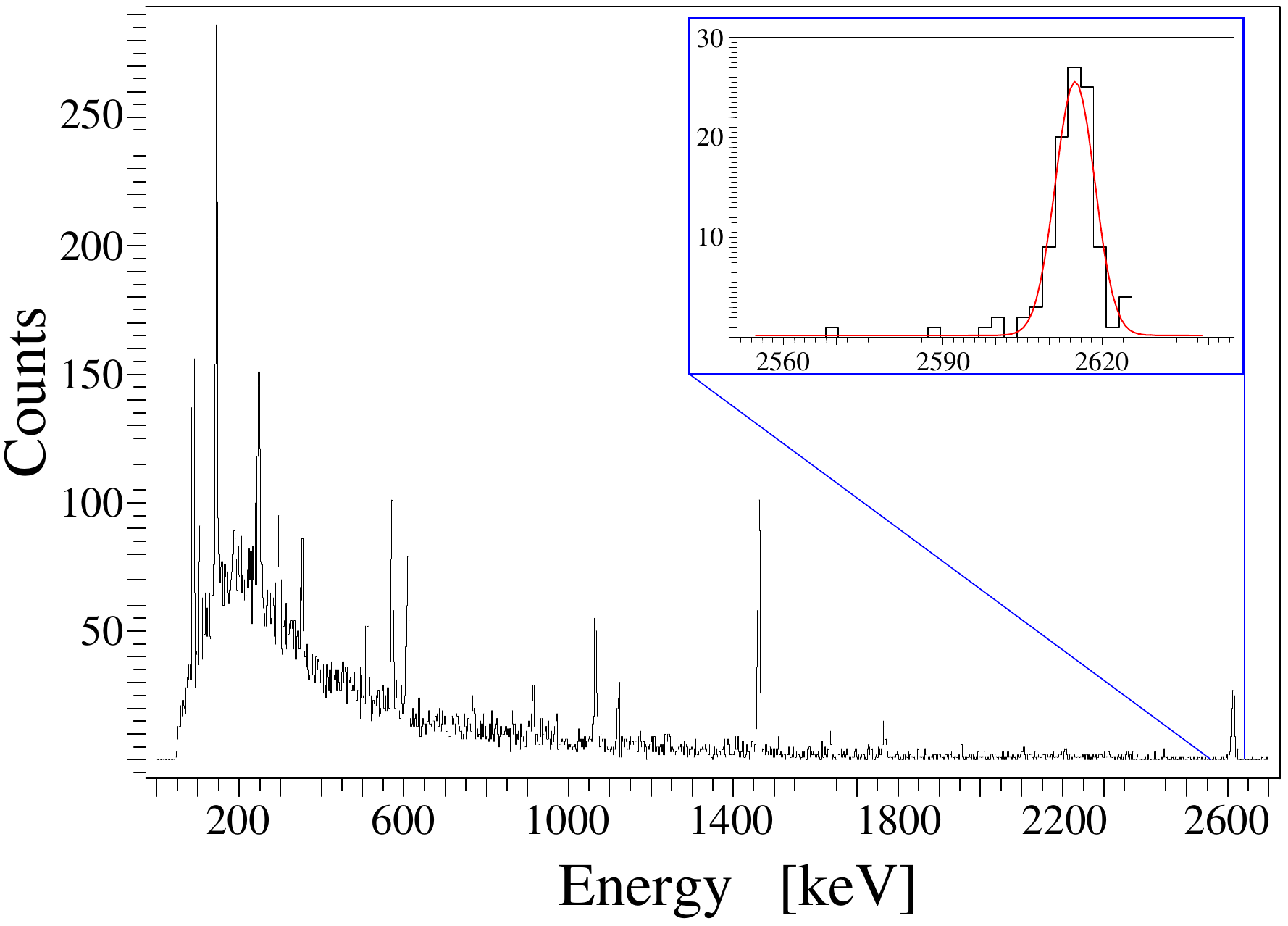}}
\caption{The 2.1 kg crystal's calibration spectrum. The most prominent low-energy lines are due to internal contaminations of $^{121m}$Te, $^{123m}$Te, 
$^{125m}$Te, $^{127m}$Te and $^{129m}$Te, activated during the shipment by airplane. Due to the high counting rate a weak $^{232}$Th 
source was used, in order to avoid pile-up. The corresponding $^{208}$Tl 2615 keV  $\gamma$-line is highlighted. 
The 1461 keV line is due to $^{40}$K environmental contamination.}
\label{fig:fig3}      
\end{center}
\end{figure}

In Fig.~\ref{fig:fig3} we present the   calibration spectrum obtained with the 2.1 kg crystal. All the observed low-energy lines are due to internal 
contaminations of metastable Te isotopes activated through  fast neutron interactions which occurred during  shipping (15 hours via airplane, from
Shanghai to Rome).
The most intense low-energy lines are 88 keV ($^{127m}$Te), 105 keV ($^{129m}$Te), 145 keV ($^{125m}$Te), 247 keV ($^{123m}$Te) and 294 keV ($^{121m}$Te).
The line at 1064 keV is due, instead, to the presence of another sample, namely a 5$\times$5$\times$5 cm$^3$ BGO 
crystal, located a few cm away in the same setup. $^{207}$Bi is a typical BGO contamination that shows a particular $\gamma$ emission at 1064 keV. 
Finally, the 1461 keV  line is due to environmental $^{40}$K contamination, while the 2615 keV line arises from the external $^{232}$Th source. The 
FWHM energy resolutions evaluated on the most intense lines are presented in Tab.~\ref{Tab:tab2}.
\begin{table}[htb]
\begin{center}
\caption{FWHM energy resolutions (in keV) evaluated from the calibration spectrum of Fig.~\ref{fig:fig1}. \dag The resolution is evaluated on 
the right (Gaussian) tail of the peak (see Fig.~\ref{fig:fig4}).}
\vspace{.2 cm}
\label{Tab:tab2}
\begin{tabular}{@{}ccccc}
\hline
145 keV  	  	& 571 keV 		   & 1461 keV   			 & 2615 keV				& 5407 keV		   	    \\
\hline
3.9 $\pm$.3	  	&  4.7 $\pm$.4	   & 6.6 $\pm$.3			 & 	7.8 $\pm$.7			& 7.8 $\pm$.2	\dag    \\
\hline
\end{tabular}
\end{center}
\end{table}

All the TeO$_2$ crystals produced so far  show a  contamination in $^{210}$Po. This contamination  does not represent a serious problem for DBD searches since it has
a ``relatively short'' decay time (T$_{1/2}$=138 days) and  is an almost pure $\alpha$-decay. Furthermore, it seems that $^{210}$Po is normally homogeneously 
distributed in the bulk so that in fact it  represents a natural calibration and stabilization line.
In this case, however, we observed a rather high contamination, of the order of 0.026 Bq. The total acquisition rate of the detector during measurement was 0.13 Hz. 
This relatively high rate (both in the $\beta$ and  $\alpha$ regions) combined with the extremely long tail of the thermal pulses results 
in a sort of ``permanent'' pile-up.
As an example, from  Fig.~\ref{fig:fig2} it can be evaluated that  $\sim$7 sec after  a  $^{210}$Po decay the baseline  is still at 1.3 \% of 
its maximum value.
Very naively one can calculate that at this point the bolometer still has to ``dissipate'' an amount of heat (i.e. energy) of roughly 
1.3 \% $\times$ 5407 keV$\approx$70 keV. This value is rather high when compared with the energy resolution presented in Tab.~\ref{Tab:tab2}.

In effect, then, iso-energy events will be randomly distributed over a decreasing tail (mostly induced by $^{210}$Po) whose amplitude 
variation can represent a large fraction with respect to the one induced by a  random event. This will affect the energy resolution.

\begin{figure}
\begin{center}
\resizebox{0.48\textwidth}{!}
{\includegraphics{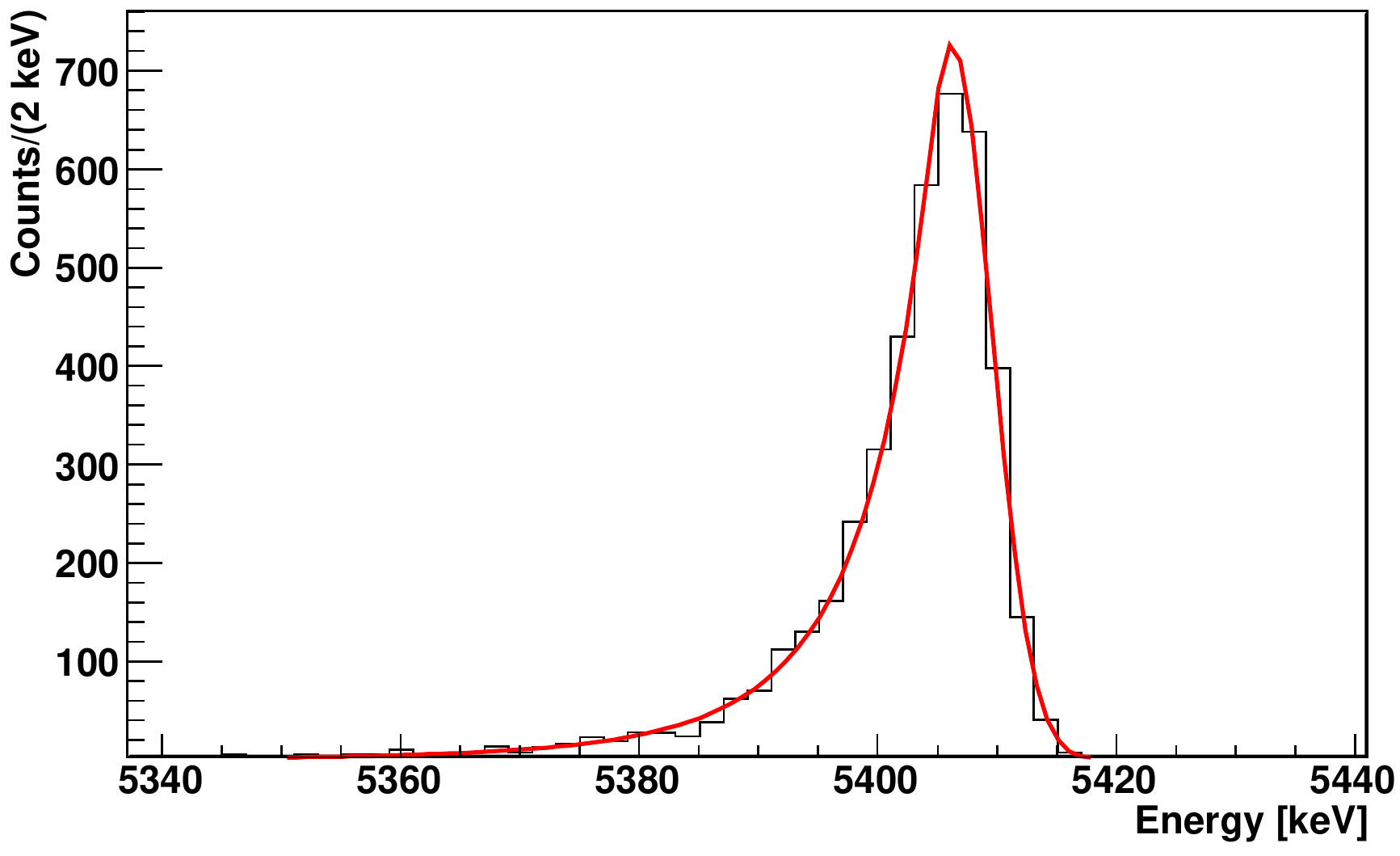}}
\caption{The $\alpha$+recoil peak due to internal contamination of $^{210}$Po. The fit, performed with the Crystal Ball function, is shown in red.}      
\label{fig:fig4} 
\end{center}
\end{figure}

The peak due to the $^{210}$Po decay is presented in Fig.~\ref{fig:fig4}.
It can be noted that the peak presents a long non-Gaussian  tail on its left side. We do not have a clear  explanation for this feature, 
even though we think it could be partially related to the high event rate of the detector. 
The tail could also be induced by  an anomalous  $^{210}$Po concentration close to the surfaces of the crystal. This is rather difficult to  evaluate
but it cannot be excluded considering  what was discussed in Sec.~\ref{sec:Growth}.
The peak is fitted with the Crystal Ball  function \cite{CBall}, which is commonly used to model various lossy processes in high-energy physics. 
The function consists of a Gaussian core portion and a power-law low-end tail, below a certain threshold.
As a final remark we would like to point out that even if the absolute signal height (24 $\mu$V/MeV) was a factor of $1/2$ less than  the expected 
one, the noise due to the electronic chain (5$\div$10 nV Hz$^{-1/2}$) is  negligible and can be estimated (for this detector) as 1.2 keV FWHM.

\section{Background considerations}
The main sources of background for TeO$_2$ bolometers \cite{fondoBB} are due to: 

\begin{itemize}
\item{$\alpha$-emitters located on or close to the surface  of a detector. }
\item{Environmental $\gamma$-emitters, mostly arising from $^{232}$Th decay chain.}
\end{itemize}

Both  contributions can be reduced by increasing the size of the crystal.

Surface contamination represents the a major source of background for the Cuoricino and CUORE experiments.
Since this background is proportional to the active surface of the crystal, while the DBD signal is proportional to the mass, decreasing the 
surface-to-volume ratio will result in an increase of the signal over background ratio.  

If for example we compare a 7.1$\times$7.1$\times$7.1 cm$^3$ TeO$_2$ crystal (which corresponds to a mass of 2.13 kg) to  a 5$\times$5$\times$5 cm$^3$ 
CUORE crystal we find the background per unit mass will be smaller by a factor  given by the ratio of their sides, 7.1/5= 1.42. 
Though this may not seem like a large number,  it must be  taken into account  that the sensitivity of a DBD experiment is directly  proportional to
(Detector Mass/Background)$^{1/2}$.

The second background effect originates from the environmental $\gamma$ radioactivity.
Before discussing the relationship between the size of the  crystal detector and the background level due to environmental $\gamma$ 
radiation, it is useful to briefly describe the main contributions to this background in the energy region  2500--2600 keV, 
since the DBD peak of $^{130}$Te is expected  at 2528 keV. 

The background in this region is largely dominated  by the $\beta$ decay of  $^{208}$Tl , belonging to the $^{232}$Th chain. 
Several high-energy $\gamma$'s are emitted within this transition, the  dominant being the 2615 keV $\gamma$-line, emitted with a Branching 
Ratio (BR) of 99\%.
Due to the extremely high transition energy of the decay (5001 keV), a cascade of other high-energy $\gamma$'s are emitted simultaneously,
the main ones being at 277 keV, 583 keV and 860 keV.

In order to understand the background, we distinguish  two different mechanisms, both involving the 2615 keV $\gamma$ line:
\begin{enumerate}
\item {multi-Compton events of the 2615 keV $\gamma$-line of $^{208}$Tl}.  

\item {multi-Compton events of the 2615 keV $\gamma$-line,  with the simultaneous interaction of  a second  $\gamma$  in the crystal}. 
\end{enumerate}
Both of these contributions vary according to the size of the detector. In the first  case, an increase in the crystal size  increases 
the probability that a $\gamma$, after several Compton interactions, releases all of its energy inside the crystal,  with a consequent \emph{improvement} 
of the peak-to-multi-Compton ratio. 

In the second case, however, with a larger crystal there is an increased  probability for two  $\gamma$'s to interact simultaneously in the crystal 
thus \emph{increasing} the background level. This contribution strongly depends on the distance between the source and the crystal.

In order to better evaluate  the background per unit mass, several Monte Carlo simulations (\textsc{Geant4}) have been run in which the distance 
between the source and  crystals of different size was varied.
In order to have enough statistics (especially for large distances) we defined the Region Of Interest (ROI) as the 
energy interval between 2500 keV and 2600 keV.
We simulated  point-like sources placed at various distances from a  5$\times$5$\times$5 cm$^3$ and a 7$\times$7$\times$7 cm$^3$ TeO$_2$ 
crystals.
In order to better understand the contribution due to the two above mentioned mechanisms, we simulated two different sources: a \emph{complete} 
$^{232}$Th decay chain, corresponding to a real physical case and a second one consisting of a single 2615 keV  $\gamma$-line emission.
In this way the contribution of the two mechanism as a  function of the distance between crystal and source can be disentangled. 

\begin{figure}
\begin{center}
\resizebox{0.48\textwidth}{!}
{\includegraphics{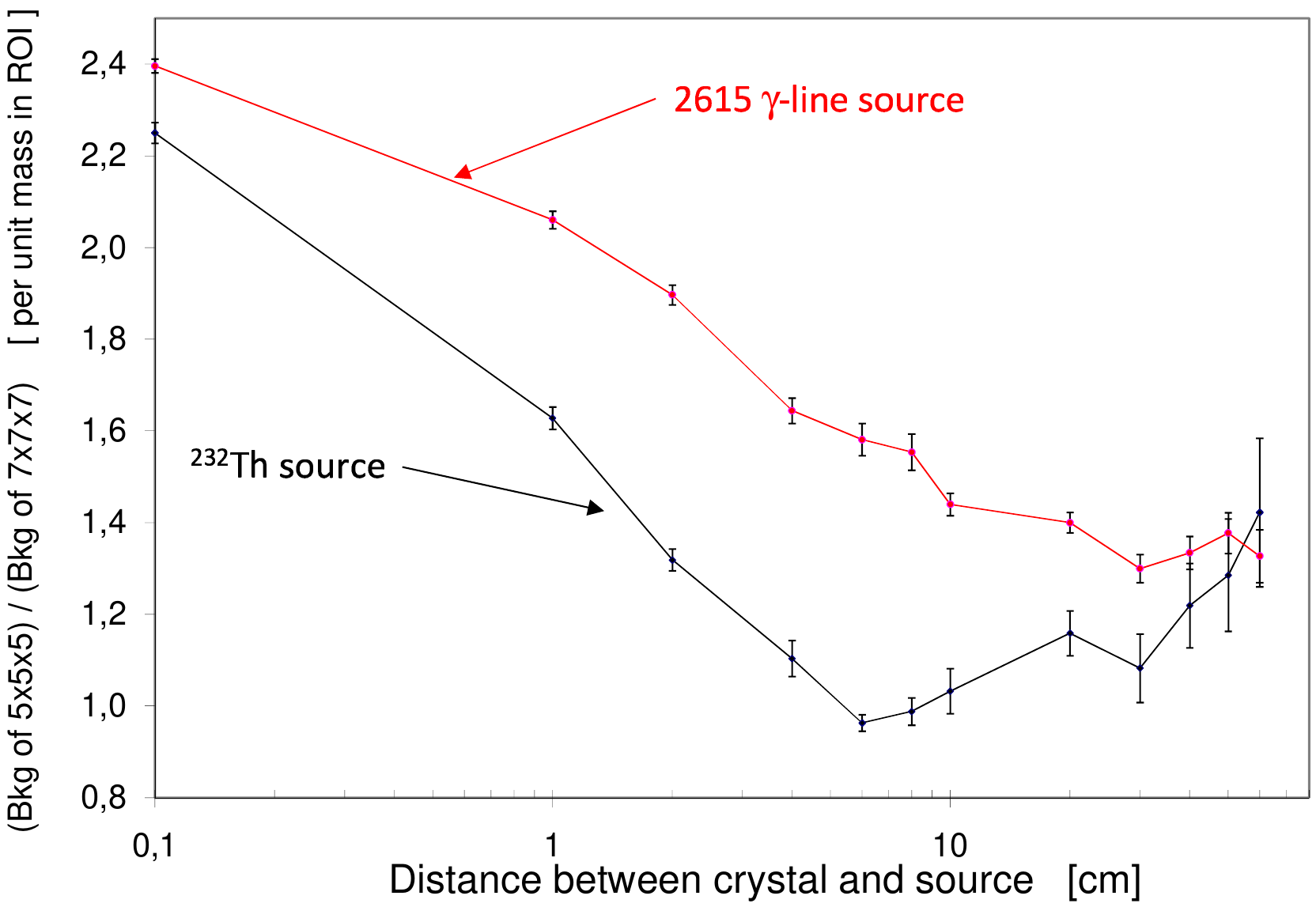}}
\caption{Ratio of the events per unit mass of a 5x5x5 cm$^3$ crystal with respect to a 7x7x7 cm$^3$ crystal evaluated in the ROI, as a function of the
distance between crystal and source.
The black points are obtained simulating a point-like ``real'' $^{232}$Th source, while the red points relate to a single 2615 keV $\gamma$-line source.
The error bars are due to statistical error and correspond to one $\sigma$-level. The lines through the data points are only to guide eyes.}      
\label{fig:fig5} 
\end{center}
\end{figure}

In Fig.~\ref{fig:fig5} we show  the results of the simulations. We plot  the ratio between the number of counts (within the ROI) per unit mass obtained with a 
5$\times$5$\times$5 cm$^3$ crystal with respect to the same for a 7$\times$7$\times$7 cm$^3$ crystal, as a 
function of the distance between the source and the crystals. 

As it can be seen, the background in the ROI for sources very close or very far from the crystal is significantly lower for larger crystals. 
For intermediate distances, were the coincidences dominate, this difference is less significant.
This can be explained in a simple way. 
When the point-like source is very close to the crystal surface, the coincidence probability does not depend strongly on crystal dimensions since it 
covers almost half of the solid angle.
Once the distance of the source from the detector increases and becomes of the same size of the crystal dimensions  then 
the \emph{difference} in the coincidence probability of the two crystals reaches its maximum, just due to a solid angle effect.
When the distance from the crystal further increases, the difference in the coincidence probability decreases again and becomes 
negligible.
We also performed simulations considering a diffused $^{232}$Th source within  a ``support material'' (such as  Cu, widely used for thermal detectors) and varying the 
distance of this support from the crystal.
Apart from the time needed for the simulation to accumulate enough events in the ROI, the ratio of the  background per unit mass is very close to the one reported
in Fig.~\ref{fig:fig5}.

We also performed some simulations considering $^{214}$Bi, the most troublesome  radionuclide of the $^{238}$U chain.
In this case we obtained only  very small differences between crystals of different size.
This is probably due to the fact that in  $^{214}$Bi $\beta$-decay there is an enormous number of $\gamma$'s emitted simultaneously.
In this case, the  term due to coincidences dominates with respect to the Multi-Compton term, resulting in a very small 
difference in the background per unit mass.
For the sake of completeness it should be remarked that for experiments based on DBD emitters whose 
transition energy does not exceed  2615 keV, the main source of environmental radioactivity is dominated by $^{232}$Th 
trace contaminations. This is simply due to the fact that the BR of  $^{214}$Bi into high energy  $\gamma$'s is 
$\sim$0.15~\% for the  $^{238}$U decay chain,  while in the case of  $^{208}$Tl the BR into the 2615 keV line is  36 \% 
for the $^{232}$Th decay chain.

We conclude by pointing out that  the efficiency of containing the 2 e$^-$ of the 0$\nu$DBD  will increase from
87.4 \% for a 0.75 kg crystal to 91.5 \% for a 2.13 kg crystal.

\section{Conclusions}
We successfully tested a 2.1 kg TeO$_2$ crystal as a thermal bolometer, the largest such detector to date.
Despite  the presence of imperfections  in the crystal, the  detector energy resolution in the  0$\nu$DBD region is the same as that obtained 
in the Cuoricino experiment.
The advantages of using larger mass crystals  for 0$\nu$DBD decay searches  was simulated and discussed.
We strongly believe that our technique is  capable of operating multi-kg crystal detectors (composed by different $\beta\beta$-emitters) 
with the required energy resolution in the 2$\div$3 MeV energy window.
There is room for improvement in crystal quality and thus in  pulse shape and energy resolution as well.
The current sample came from a  routine production, so better-quality large crystals could be obtained if dedicated furnaces were used.

\section {Acknowledgements}
Thanks are due to the LNGS mechanical workshop, in particular to E. Tatananni, A. Rotilio, A. Corsi, B. Romualdi and F. De Amicis  for their continuous 
and constructive help in each aspect of detector design and construction. 
We are especially grateful to Maurizio Perego for his invaluable help in the development and 
improvement  of the data acquisition software.

\end{document}